\documentclass[nofootinbib,aps,prc]{revtex4-1}

\usepackage{epsfig}
\usepackage[textwidth=15.5cm]{geometry}
\usepackage[utf8]{inputenc}
\usepackage{amsmath}
\usepackage{amssymb}
\usepackage{amsthm}
\usepackage{amsfonts}
\usepackage{xspace}
\usepackage{graphicx}
\usepackage{graphics}
\usepackage{slashed}
\usepackage{bbold}

\usepackage{cancel}

\usepackage{hyperref}
\hypersetup{
    colorlinks,
    linkcolor={red!50!black},
    citecolor={blue!50!black},
    urlcolor={blue!80!black}
}

\newcommand{\Nc}{\ensuremath{N_c}\xspace}
\newcommand{\LONc}{\ensuremath{\text{LO-in-\Nc}}\xspace}
\newcommand{\NNLONc}{\ensuremath{\text{N${}^2$LO-in-\Nc}}\xspace}

\newcommand{\oneS}{{{}^{1}\!S_0}}
\newcommand{\threeS}{{{}^{3}\!S_1}}
\newcommand{\onePone}{{{}^{1}\!P_1}}
\newcommand{\Ptwo}{{{}^{3}\!P_2}}
\newcommand{\Pone}{{{}^{3}\!P_1}}
\newcommand{\Pzero}{{{}^{3}\!P_0}}
\newcommand{\PJ}{{{}^{3}\!P_J}}
\newcommand{\oneP}{{{}^{1}\!P_1}}

\newcommand{\LRd}{\overset{\leftrightarrow}{\nabla}{}}
\newcommand{\LRC}{\overset{\leftrightarrow}{C}{}}

\newcommand{\calO}{\ensuremath{\mathcal{O}}}

\newcommand{\calL}{\ensuremath{\mathcal{L}}}

\renewcommand\vector{\mathbf}

\newcommand{\pplus}{\vector{p}_+}
\newcommand{\pminus}{\vector{p}_-}

\newcommand{\Czerotrip}{\ensuremath{C_0^{(^3 \! S_1)}}\xspace}
\newcommand{\Czerosing}{\ensuremath{C_0^{(^1 \! S_0)}}\xspace}
\newcommand{\Ctrip}{\ensuremath{C_2^{(^3 \! S_1)}}\xspace}
\newcommand{\Csing}{\ensuremath{C_2^{(^1 \! S_0)}}\xspace}
\newcommand{\Csd}{\ensuremath{C^{(SD)}}\xspace}
\newcommand{\ConeP}{\ensuremath{C^{(^1 \! P_1)}}\xspace}
\newcommand{\CPzero}{\ensuremath{C^{(^3 \! P_0)}}\xspace}
\newcommand{\CPone}{\ensuremath{C^{(^3 \! P_1)}}\xspace}
\newcommand{\CPtwo}{\ensuremath{C^{(^3 \! P_2)}}\xspace}

\newcommand{\asing}{\ensuremath{a^{(^1 \! S_0)}}\xspace}
\newcommand{\rsing}{\ensuremath{r^{(^1 \! S_0)}}\xspace}
\newcommand{\atrip}{\ensuremath{a^{(^3 \! S_1)}}\xspace}
\newcommand{\rtrip}{\ensuremath{r^{(^3 \! S_1)}}\xspace}

\newcommand{\mpi}{\ensuremath{m_\pi}}

\newcommand{\eftnopi}{EFT$_{\pi\hskip-0.40em /}$\xspace}
\newcommand{\NXLO}[1]{N\ensuremath{{}^{#1}}LO\xspace}
\newcommand{\NtwoLO}{\NXLO{2}}

\newcommand{\LambdaNoPion}{\ensuremath{\Lambda_{\pi\hskip-0.4em /}}}

\newcommand{\Ndag}{N^\dagger}

\usepackage{siunitx}
\usepackage{xcolor}

\usepackage{bm,upgreek}

\newcommand\SD{$S$-$D$}

\newcommand\couplingOneZero{C_{1\cdot 1}}
\newcommand\couplingThreeOne{C_{G\cdot G}}
\newcommand\couplingSixOne{C_{G \cdot G}'}
\newcommand\couplingOneOne{C_{\tau \cdot \tau}}
\newcommand\couplingTwoZero{{\LRC}_{1 \cdot 1}}
\newcommand\couplingThreeZero{C_{\sigma\cdot \sigma}}
\newcommand\couplingFourOne{{\LRC}_{G\cdot G}}
\newcommand\couplingFiveZero{{\LRC}_{1\cdot \sigma}}
\newcommand\couplingFiveOne{{\LRC}_{G\cdot \tau}}
\newcommand\couplingSixZero{C_{\sigma\cdot\sigma}'}
\newcommand\couplingSevenOne{\LRC^\prime_{G \cdot G}}

\begin{document}

\title{Large-\Nc Relationships Among Two-Derivative Pionless Effective Field Theory Couplings}

\author{Matthias R.~Schindler}
\email{mschindl@mailbox.sc.edu}
\affiliation{Department of Physics and Astronomy,\protect\\
University of South Carolina,\protect\\
Columbia, SC 29208}

\author{Hersh Singh}
\email{hersh@phy.duke.edu}
\affiliation{Department of Physics,\protect\\
Duke University,\protect\\
Durham, NC 27708}

\author{Roxanne P.~Springer}
\email{rps@phy.duke.edu}
\affiliation{Department of Physics,\protect\\
Duke University,\protect\\
Durham, NC 27708}

\date{\today}

\begin{abstract}

We analyze two-derivative two-nucleon interactions in a combined pionless effective field theory and large-\Nc expansion.  At leading order in the large-\Nc expansion, relationships among low-energy constants emerge. We find these to be consistent with experiment.   However, it is critical to correctly address the subtraction-point dependence of the low-energy constants.  These results provide additional confidence that the dual-expansion procedure is useful for analyzing low-energy few-body observables.  

\end{abstract}

\maketitle


\section{Introduction}

In the effective field theory (EFT) approach to nucleon-nucleon (NN) interactions, the underlying short-distance details of quantum chromodynamics (QCD) are encoded in the values of the low-energy couplings (LECs) that accompany each operator. 
At energies well below the pion mass, the nucleon-nucleon interactions are given  in terms of two-nucleon contact operators with an increasing number of derivatives as well as any needed external fields. This approach is commonly referred to as pionless EFT (\eftnopi), see, e.g., Refs.~\cite{Chen:1999tn,vanKolck:1999mw,Beane:2000fx,Bedaque:2002mn,Platter:2009gz} and references therein. 
While in the future it may be possible to predict the LECs from QCD directly, currently their values are determined from fits to experimental observables or phase shift analyses of data. 
As we attempt to reconcile increasingly accurate low-energy measurements with increasingly accurate calculations, the number of LECs needed in \eftnopi increases.
Additional theoretical constraints can provide relationships between the LECs and reduce the number of independent couplings.

Here we analyze the two-derivative nucleon-nucleon contact operators contributing to low-energy scattering by considering the  large-\Nc   expansion of QCD, where \Nc is the number of colors \cite{tHooft:1973jz,Witten:1979kh}. 
In the large-\Nc limit additional symmetries emerge. Fortunately, in many instances these symmetries survive as approximate symmetries in actual ($\Nc= 3$) QCD. Combining the \eftnopi and large-\Nc expansions decreases the number of independent LECs and increases the predictive power of the combined expansions.
For example, when calculating an observable for which insufficient data is available to fit the higher-order (in the \eftnopi expansion) LECs, the large-\Nc relationships may make predictions possible.

Large-\Nc methods were applied to nucleon interactions in Ref.~\cite{Kaplan:1995yg}, which showed that in the large-\Nc limit the nonderivative $S$-wave interactions should be identical. The general two-nucleon potential was analyzed in Ref.~\cite{Kaplan:1996rk}, predicting the relative strength of central, tensor, and spin-orbit forces. These results compared favorably with the couplings as predicted by the Nijmegen potential model \cite{Stoks:1994wp} derived from an analysis of a large set of scattering data over a wide range of energies. Additional phenomenological models were considered in Ref.~\cite{Riska:2002vn}. 
Reference \cite{CalleCordon:2008cz,*CalleCordon:2009ps,*RuizArriola:2016vap} investigated Wigner-SU(4) and Serber symmetries, and explored their connection to large-\Nc in NN potential models.
The large-\Nc approach has also been applied to three-nucleon interactions \cite{Phillips:2013rsa,Epelbaum:2014sea}.

EFTs have emerged as powerful tools for describing nucleon interactions (see, e.g., Refs.~\cite{vanKolck:1999mw,Bedaque:2002mn,Epelbaum:2008ga,Platter:2009gz,Machleidt:2011zz} and references therein).
In Ref.~\cite{Schindler:2015nga} some of us applied the dual \eftnopi and large-\Nc expansions to low-energy two-nucleon parity-violating interactions.  Since there is not enough data to confirm (although current data is not in contradiction with) those results, we want to test the formalism on a system where more data are available.
Here, in addition to the large-\Nc expansion, we focus on low energies and parity-conserving interactions in partial waves with orbital angular momentum $L\le 2$ as described by \eftnopi. In  this formalism, these interactions are described by nucleon contact operators with up to two derivatives. 
The two-derivative two-nucleon operators are interesting for several reasons. First, the lowest-order (no-derivative) \eftnopi operators (in the $\oneS$  and  $\threeS$ channels) can obscure the large-\Nc{} relationships if not analyzed carefully. 
The large-\Nc prediction that the LECs in these two channels are the same at leading order in \Nc (\LONc) \cite{Kaplan:1995yg}  suggests that the $S$-wave scattering lengths in the large-\Nc limit should be identical. In reality, the scattering lengths are not close to one another and even differ in sign, which seems to contradict the large-\Nc prediction. However, the $S$-wave scattering lengths are known to be fine-tuned and anomalously large (see also Ref.~\cite{Mehen:1999qs}). 
Further, the LECs themselves, to which the large-\Nc predictions apply, are not observables and are renormalization scheme dependent. The values of the LECs agree within the expected accuracy once an appropriate renormalization point is chosen.

A more robust test of the large-\Nc{}  relationships  may be found by considering two-nucleon scattering at higher orders in \eftnopi because the effective range expansion parameters at these orders (the effective range, shape parameter,  etc.) are expected to be of natural size.
The two-body $P$-wave and \SD{}-mixing operators, which first occur at two derivatives, are important for the role they play in currently unexplained asymmetries in three-body scattering at very low energies ($E_\text{nucleon} < \SI{12}{MeV}$).
The suite of polarization observables in nucleon-deuteron scattering depends upon both two-nucleon \SD{} mixing and two-body scattering in the $^{3}\!P_{J}$, $J=0,1,2$ channels. In particular, the so-called $A_y$  (nucleon-deuteron) analyzing power is very sensitive to two-body scattering in the ${{}^{3}\!P_{J}}$ channels.
When the ${}^{3}\!P_{J}$ phase shifts are modified by only 15\% it can yield changes in $A_y$ by 50\%, which easily accounts for the current $A_y$ discrepancies between theory and experiment. This issue is reviewed in depth in Ref.~\cite{Gloeckle:1995jg} and references therein and considered in \eftnopi in Ref.~\cite{Margaryan:2015rzg}.  

In the next section we will present the needed terminology for defining and using \eftnopi and large-\Nc techniques.


\section{Pionless EFT and large-\texorpdfstring{\Nc}{Nc} constraints}

\eftnopi describes nucleon-nucleon interactions at energies of less than tens of \si{MeV} in terms of contact terms that only involve nucleon fields and external currents. These operators are organized according to a power counting based on a small expansion parameter $Q/\LambdaNoPion$. All short-distance/high-energy details are encoded in the corresponding LECs. 
Because of the number of LECs and the sparse data available, some with potentially large error bars at the lowest scattering energies, we will explore the enhanced symmetry that QCD acquires in the limit \Nc $\rightarrow \infty$.  This symmetry is approximate for real ($\Nc= 3$) QCD, but corrections are expected to be perturbative in 1/\Nc.   Combining the \eftnopi and large-\Nc expansions allows us to reduce the number of LECs occurring at any given order  compared to the standard \eftnopi expansion.

The \eftnopi expansion parameter is $Q/\LambdaNoPion$, where $Q$ is the momentum and/or energy transfer and $\LambdaNoPion \sim \mpi$ is the breakdown scale of the theory. 
However, a  naive counting of derivatives---as is possible in, for example, chiral perturbation theory---is not sufficient to establish $Q$ scaling in \eftnopi.  In particular,  the large scattering lengths in the $S$-wave channels for two-nucleon  scattering require a modification of this naive scaling \cite{Kaplan:1996xu,Kaplan:1998we,Kaplan:1998tg,vanKolck:1998bw}.  The power divergent subtraction (PDS) scheme \cite{Kaplan:1998tg}  introduces a scale  (subtraction point) $\mu \sim Q$ in the problem to establish a consistent power counting in the anomalous $S$-wave channels. 
For example, the leading (no-derivative) $S$-wave interactions are given by
\begin{align}
\label{eq:LOSwave}
  \mathcal{L}_{0}^{(S)}=-C_{0}^{({}^{3}\!S_{1})}(N^{T} P_i N)^{\dagger}(N^{T}P_i N) 
-C_{0}^{({}^{1}\!S_{0})}( N^{T} P_{a} N)^{\dagger}(N^{T}P_{a} N) \ ,
\end{align}
where $P_i=\frac{1}{\sqrt{8}}\sigma_{2}\sigma_{i}\tau_{2}$ and $P_a=\frac{1}{\sqrt{8}}\sigma_{2} \tau_{2}\tau_{a}$,
with $\sigma_i$ $(i=1,2,3)$ and $\tau_a$ $(a=1,2,3)$  the Pauli matrices in spin and isospin space.
The operators are dimension 6, so their coefficients have units of \si{MeV^{-2}}.
However,  these two energy dimensions are not both from large scales in the problem, as is typical in many EFTs.  Instead, in the PDS scheme, the coefficients scale as 
\begin{equation}
\label{leadingCs}
  C_{0}^{(S)}=\frac{4 \pi}{M}\frac{1}{\frac{1}{a^{(S)}}-\mu} \ , 
\end{equation}
where $M$ is the nucleon mass, $\mu$ is the subtraction scale in the PDS scheme, and $a^{(S)}$ is the scattering length in the $S$ channel, either $\threeS$ or $\oneS$. 
For $\mu\sim Q$, this yields an $S$-wave scattering amplitude of the order $1/(MQ)$.  As we will see below, the modification of naive power counting in the $S$-wave channels will impact the \eftnopi $Q$ scaling of higher-order $S$-wave coefficients as well.
But see Sec.~\ref{subtraction} for a discussion on restrictions on $\mu$ when analyzing $\mu$-dependent (non-physical) quantities.

The \eftnopi operators of Eq.~\eqref{eq:LOSwave} are given in the so-called partial-wave basis, where the two incoming and the two outgoing nucleons are identified by their $^{2S+1}L_J$ quantum numbers, where $S=0,1$ is the total intrinsic spin, $L$ is the relative orbital angular momentum of the two nucleons, and $J$ is the total angular momentum  of the two-nucleon system. The $J$ quantum numbers are conserved but others may mix. 
This basis is particularly convenient because it maps most directly onto experimental results.  

However, the partial-wave basis is not the basis in which large-\Nc counting is clearly manifest. The large-\Nc counting of two-nucleon operators is determined by considering the  two-nucleon matrix element of the Hamiltonian $H$ \cite{Kaplan:1996rk},
\begin{equation}\label{potential}
V(\pminus,\pplus) = \langle (\vector{p}_1^\prime, C),(\vector{p}_2^\prime, D) \vert H \vert (\vector{p}_1, A),(\vector{p}_2, B) \rangle  \ ,
\end{equation}
where the $A,B,C,D$ denote combined spin and isospin quantum numbers of the nucleons and 
\begin{equation}
\vector{p}_\pm \equiv \vector{p}^\prime \pm \vector{p}  \ ,
\end{equation}
with the outgoing and incoming relative momenta given by
\begin{equation}
\vector{p}^\prime = \vector{p}^\prime_1 - \vector{p}^\prime_2 \ , \quad \vector{p} = \vector{p}_1 - \vector{p}_2  \ ,
\end{equation}
in terms of the momenta in Eq.~\eqref{potential}.
$H$ is the Hartree Hamiltonian \cite{Witten:1979kh,Kaplan:1996rk},
\begin{equation}
H = \Nc \sum_n\sum_{s,t} v_{stn} \left(\frac{S}{\Nc}\right)^s \left(\frac{I}{\Nc}\right)^t \left(\frac{G}{\Nc}\right)^{n-s-t} \ ,
\end{equation}
expressed in terms of the operators
\begin{equation}
\label{eq:q-ops}
S_i = q^\dagger \frac{\sigma_i} {2} q \ , \quad I_a=q^\dagger \frac{\tau_a} {2}q \ , \quad G_{ia}=q^\dagger \frac{\sigma_i \tau_a} {2}q  \ ,
\end{equation}
where $q$ are bosonic (colorless) SU(2) doublet operators. 
The coefficients $v_{stn}$ account for momentum dependence and the desired symmetry properties.

The large-\Nc counting of matrix elements was explored in Refs.~\cite{Dashen:1993jt,Dashen:1994qi,Kaplan:1995yg,Kaplan:1996rk}, which found scaling of the form
\begin{equation}
\label{eq:op-scale}
 \langle N(\vector{p}^\prime) | \frac{\mathcal{O}_{I,S}^{(n)}}{\Nc^n} | N(\vector{p})\rangle \lesssim \frac{1}{\Nc^{|I-S|}}  \ ,
 \end{equation}
where $\mathcal{O}_{I,S}^{(n)}$ is an $n$-body quark operator of isospin $I$ and spin $S$ (e.g., $q^\dagger q$ is an  $n$=1 one-body  operator). 
Using Eq.~\eqref{eq:op-scale} on the quark-model operators of Eq.~\eqref{eq:q-ops}, the large-\Nc scaling of these operators is
\begin{equation}
\langle N(\vector p^\prime) | \frac{S}{\Nc} | N(\vector p)\rangle \sim \langle N(\vector p^\prime) | \frac{I}{\Nc} | N(\vector p)\rangle \lesssim \frac{1}{\Nc} \ ,  \quad \langle N(\vector p^\prime) | \frac{G}{\Nc} | N(\vector p)\rangle \lesssim 1 \ .
\end{equation}
In addition, the identity operator scales as $\langle N(\vector p^\prime) | \mathbb{1} | N(\vector p)\rangle \sim \Nc$.

In principle, the momenta $\pminus$ and $\pplus$ are both considered to be independent of \Nc. However, as discussed in Ref.~\cite{Kaplan:1996rk}, there is a possible additional 1/\Nc suppression related to these momenta. In the large-\Nc limit, it is consistent to interpret the potential in the meson-exchange picture \cite{Banerjee:2001js,Cohen:2002im}. In the $t$-channel, contributions proportional to $\pplus$ only arise as relativistic corrections and are always accompanied by a factor of $1/M\sim 1/\Nc$. Analogously, in the $u$-channel $\pminus$ only appears as a relativistic correction with a factor of $1/M$. Instead of considering all possible contractions of the operators between two-nucleon states, it is convenient to only consider $t$-channel contributions, for which the large-\Nc suppression of relativistic corrections can be taken into account by counting the momenta as 
\begin{align}
\pminus \sim 1 \ , \quad \pplus \sim 1/\Nc \ .
\end{align}
Analysis of the $u$-channel contributions would lead to equivalent results, see also the discussion in Ref.~\cite{Phillips:2013rsa}.

In the large-\Nc limit the matrix element of Eq.~\eqref{potential} factorizes into matrix elements of single-nucleon operators; for our two-body-scattering purposes we perform the large-\Nc analysis on operators of the form $(N^\dagger \mathcal{O}_1 N)(N^\dagger \mathcal{O}_2 N)$, where the $\mathcal{O}_i$ are spin-isospin operators. The large-\Nc scaling found from these matrix elements can be mapped onto the \eftnopi LECs. By using Fierz transformations, the large-\Nc behavior of the LECs in the \eftnopi partial-wave basis can then be extracted.

One issue that remains an open question in this procedure is the role of the $\Delta$ resonance in the large-\Nc limit of QCD.  From an EFT viewpoint we are free to integrate out the $\Delta$ as well as the pion to form a low-energy theory.  But the large-\Nc limit does not necessarily respect that point of view.  In particular, in the large-\Nc limit the nucleon-$\Delta$ mass splitting goes to zero and it is reasonable to be concerned about whether one can legitimately talk about the large-\Nc limit in \eftnopi without the $\Delta$ degree of freedom. 
In the following we will restrict the discussion to the matrix element defined in Eq.~\eqref{potential} without considering the effects of virtual $\Delta$s in intermediate states. A more detailed discussion about the impact of integrating out  $\Delta$ degrees of freedom can be found in \cite{Savage:1996tb}.

In the next section we consider the two-derivative \eftnopi operators needed to describe $S$-wave, $P$-wave, and \SD{}-mixed  two-nucleon scattering.


\section{Two-derivative operators}
\label{sec-twoderivative}

In \eftnopi, there are seven independent two-nucleon operators with two derivatives. In the partial-wave basis, these correspond to the two $S$-wave, the \SD{}-mixing, and the four $P$-wave terms. In what we will call the ``large-\Nc-counting basis,'' there appears to be a great deal of freedom in choosing a set of seven independent terms to take as the basis.   In the absence of large-\Nc considerations, Fierz relationships  show the equivalence of these choices.
However, as discussed in Ref.~\cite{Schindler:2015nga}, the large-\Nc counting is most transparent when performed on an overcomplete set of the large-\Nc-basis operators, because the Fierz transformations that are used to reduce the Lagrangian to a minimal  set can relate terms that are of the same order in the \eftnopi counting, but have different large-\Nc scalings.

The overcomplete basis required for our analysis contains a total of 14 terms \cite{Ordonez:1995rz,Epelbaum:1998ka,Girlanda:2010ya}. Since we are interested in isospin-singlet terms, each operator of a given spin-momentum structure appears in two possible forms: either with the  isospin identity operators, or with $\boldsymbol{\tau}_1 \cdot \boldsymbol{\tau}_2$.

The two-derivative operators of the form $(N^\dagger \mathcal{O}_1 N)(N^\dagger \mathcal{O}_2 N)$  that are \LONc  are $O(\Nc)$ and can be chosen to be 
\begin{align}\label{LONc}
  \calL_\text{\LONc} &=  \couplingOneZero \nabla_i (\Ndag N )  \nabla_i (\Ndag N )  \nonumber \\
                  &\quad  + \couplingThreeOne \nabla_i (\Ndag \sigma_j \tau_a N )  \nabla_i (\Ndag \sigma_j \tau_a N  )  \\
                  & \quad + \couplingSixOne \nabla_i (\Ndag \sigma_i \tau_a N )  \nabla_j (\Ndag \sigma_j \tau_a N  ) \ . \nonumber
\end{align}
The subscripts on the coefficients are designed to echo the operators they multiply.  When more than one contraction is possible a prime superscript is used.

Given that the expansion parameter is $1/\Nc=1/3$ for real QCD, corrections to these leading (in $\Nc$) terms might be expected at the 30\% level.
However, the next terms in the large-\Nc expansion for these processes occurs two orders higher, $O(1/\Nc)$.  Those operators may be chosen to be
\begin{align}\label{NLONc}
  \calL_\text{\NNLONc} & = \couplingOneOne \nabla_i (\Ndag \tau_a N )  \nabla_i (\Ndag \tau_a N ) \nonumber \\
                   & \quad + \couplingTwoZero (\Ndag \LRd_i N )  (\Ndag \LRd_i N ) \nonumber \\
                   & \quad + \couplingThreeZero \nabla_i (\Ndag \sigma_j  N )  \nabla_i (\Ndag \sigma_j N  ) \nonumber \\
                   & \quad + \couplingFourOne (\Ndag \LRd_i\sigma_j \tau_a N )  (\Ndag \LRd_i \sigma_j \tau_a N  )   \\
                   & \quad - \frac{i}{2}\couplingFiveZero \ \epsilon_{ijk} \left[ \nabla_j (\Ndag \sigma_i  N )  (\Ndag \LRd_k  N) + \nabla_j (\Ndag  N )  (\Ndag \LRd_k \sigma_i N  ) \right] \nonumber \\
                   & \quad - \frac{i}{2}\couplingFiveOne \ \epsilon_{ijk} \left[ \nabla_j (\Ndag \sigma_i \tau_a N )  (\Ndag \LRd_k \tau_a N) + \nabla_j (\Ndag  \tau_a N )  (\Ndag \LRd_k \sigma_i \tau_a N  ) \right] \nonumber \\
                   & \quad + \couplingSixZero \nabla_i (\Ndag \sigma_i  N )  \nabla_j (\Ndag \sigma_j  N  ) \nonumber \\
                   & \quad + \couplingSevenOne (\Ndag \LRd_i\sigma_i \tau_a N )  (\Ndag \LRd_j \sigma_j \tau_a N  ) \ , \nonumber 
\end{align}
where \NtwoLO denotes next-to-next-to-leading order. The remaining three operators from the initial overcomplete set of 14 are suppressed by additional powers of 1/\Nc and will not be considered here.

As discussed above, only 7 of the 14 operator structures are independent. There is some freedom in choosing which set of operators to retain. In EFT calculations in this basis, Fierz transformations can be applied to remove the contributions containing $\boldsymbol{\tau}_1 \cdot \boldsymbol{\tau}_2$ in favor of the simpler identity operators.
However, this approach masks the large-\Nc counting, since eliminating the terms proportional to $\couplingThreeOne$ and $\couplingSixOne$ induces LO-in-\Nc contributions in terms that are naively of higher order in \Nc. It is possible but tedious to keep track of these ``hidden'' LO-in-\Nc contributions. Instead, it is more convenient to keep all terms in Eq.~\eqref{LONc} and to apply Fierz transformations to the \NtwoLO-in-\Nc terms in Eq.~\eqref{NLONc}.  Four of these terms are redundant and can be removed. We choose to keep the terms proportional to $\couplingOneOne$, $\couplingThreeZero$, $\couplingFiveZero$, and  $\couplingSixZero$.

We wish to use the large-\Nc scaling of the above operators to indicate how the partial-wave operators scale with large-\Nc.  The two-derivative \eftnopi operators in the partial-wave basis are 
\begin{align}
  \calL_2^{(^3 \! S_1)} & = \frac{1}{8} \Ctrip \left[ (N^T P_i N)^\dagger (N^T P_i \LRd^2 N) + \text{h.c.}\right] \ , \\
  \calL_2^{(^1 \! S_0)} & =  \frac{1}{8} \Csing \left[ (N^T P_a N)^\dagger (N^T P_a \LRd^2 N) + \text{h.c.}\right]  \ , \\
  \calL _2^{(SD)}& =  \frac{1}{4} \Csd \left[ (N^T P_i N)^\dagger (N^T P_j \LRd_x \LRd_y N)(\delta_{ix}\delta_{jy} - \frac{1}{3}\delta_{ij}\delta_{xy}) + \text{h.c.}\right]  \ , \\
  \calL _2^{(^1 \! P_1)}& =  \frac{1}{4} \ConeP (N^T P_0 \LRd_i  N)^\dagger (N^T P_0 \LRd_i  N) \label{1p1} \ , \\
  \calL _2^{(^3 \! P_J)}& =  \frac{1}{4}  \left[  \CPzero \delta_{xy}\delta_{wz}+ \CPone (\delta_{xw}\delta_{yz}-\delta_{xz}\delta_{yw}) \right. \label{3pj} \nonumber\\                        
  & \quad \left. + \CPtwo (2\delta_{xw}\delta_{yz}+2\delta_{xz}\delta_{yw}-\frac{4}{3}\delta_{xy}\delta_{wz}) \right]  (N^T P_{y,a} \LRd_x  N)^\dagger (N^T P_{z,a} \LRd_w  N) \ , 
\end{align}  
where $N^T \calO \LRd_i N \equiv N^T \calO \nabla_i N - (\nabla_i N^T) \calO N$  with $\calO$ some spin-isospin operator, and the projection operators are defined by 
\begin{align} \label{proj}
  P_i & = \frac{1}{\sqrt{8}} \sigma_2 \sigma_i \tau_2 \ , 
  & P_a & = \frac{1}{\sqrt{8}} \sigma_2  \tau_2 \tau_a \ , 
  & P_0 & = \frac{1}{\sqrt{8}} \sigma_2 \tau_2 \ , 
  & P_{i,a} & = \frac{1}{\sqrt{8}} \sigma_2 \sigma_i \tau_2 \tau_a \ .
\end{align}

The Fierz identities in Appendix~\ref{app} can be used to relate the large-\Nc-counting basis to the partial-wave basis. The \eftnopi partial-wave coefficients have large-\Nc scaling provided by
\begin{equation}
  C^{(^{2S+1}L_J)}= C^{(^{2S+1}L_J)}_\text{\LONc}+C^{(^{2S+1}L_J)}_\text{\NNLONc}  \ ,
\end{equation}
where each $C^{(^{2S+1}L_J)}_\text{\LONc}$ is a linear combination of $\couplingOneZero$, $\couplingThreeOne$, and $\couplingSixOne$ and each $C^{(^{2S+1}L_J)}_\text{\NNLONc}$ is a linear combination of $\couplingOneOne$, $\couplingThreeZero$, $\couplingFiveZero$, and $\couplingSixZero$:
\begin{align}\label{NLOrelns}
  \Csing  = &-4 (\couplingOneZero - 3 \couplingThreeOne - \couplingSixOne) \ \ &\text{\LONc} \nonumber  \\ 
  & - 4(\couplingOneOne- 3\couplingThreeZero  - \couplingSixZero )  \ \ &\text{\NNLONc} \nonumber \\
  \Ctrip   = & -4 (\couplingOneZero - 3 \couplingThreeOne - \couplingSixOne) \ \ &\text{\LONc} \nonumber  \\ 
  & -4( -3 \couplingOneOne + \couplingThreeZero  + \frac{1}{3}\couplingSixZero ) \ \ &\text{\NNLONc} \nonumber \\ 
  \Csd  =& - 4 ( 3\couplingSixOne) &\text{\LONc} \nonumber \\
  &-4( - \couplingSixZero)&\text{\NNLONc} \nonumber \\
  \ConeP  = & -4 (\couplingOneZero + 9 \couplingThreeOne + 3 \couplingSixOne) &\text{\LONc}\nonumber \\
  &-4( - 3 \couplingOneOne - 3 \couplingThreeZero  - \couplingSixZero )&\text{\NNLONc} \nonumber \\
  \CPzero  = & -\frac{4}{3} (\couplingOneZero + \couplingThreeOne - 3 \couplingSixOne) &\text{\LONc} \nonumber \\
   &-\frac{4}{3}( \couplingOneOne  + \couplingThreeZero  - 2 \couplingFiveZero  - 3 \couplingSixZero) &\text{\NNLONc} \nonumber \\
  \CPone  = &-2 (\couplingOneZero + \couplingThreeOne + 2 \couplingSixOne) &\text{\LONc} \nonumber \\
  &-2(  \couplingOneOne  + \couplingThreeZero - \couplingFiveZero  + 2 \couplingSixZero) &\text{\NNLONc} \nonumber \\
  \CPtwo  = &-(\couplingOneZero +  \couplingThreeOne) &\text{\LONc}\nonumber \\
  &-(  \couplingOneOne + \couplingThreeZero  + \couplingFiveZero) \ . &\text{\NNLONc} 
\end{align}              

While the large-\Nc basis of Eqs.~(\ref{LONc}) and (\ref{NLONc}) is the one where large-\Nc scaling is manifest, and the partial-wave basis has obvious physical significance, there is another basis that is physically illuminating: the one that is driven by whether the operator is of central, tensor, or spin-orbit type.  From inspection we can see that the leading-order \eftnopi terms up to \NtwoLO-in-\Nc counting can be categorized as central (derivatives contract with themselves): $\couplingOneZero$, $\couplingThreeOne$, $\couplingOneOne$,  and $\couplingThreeZero$;  tensor (derivative contracts with $\vec \sigma$): $\couplingSixOne$ and $\couplingSixZero$; and spin-orbit (cross product): $\couplingFiveZero$.  The contribution of central vs.~tensor vs.~spin-orbit to partial-wave channels is known, as is their large-\Nc counting from Ref.~\cite{Kaplan:1996rk}, see, for example, Ref.~\cite{CalleCordon:2008cz,*CalleCordon:2009ps,*RuizArriola:2016vap}; here we delineate them in terms of \eftnopi counting as well. The tensor interaction also provides a clear example of the necessity to consider the overcomplete Lagrangian for the large-\Nc analysis. As shown in Ref.~\cite{Kaplan:1996rk} and in agreement with our calculation, the tensor interaction is of LO in \Nc. However, it is the operator $\couplingSixOne$  proportional to $\boldsymbol{\tau}_1 \cdot \boldsymbol{\tau}_2$ that gives the LO contribution. Had we eliminated this term using Fierz identities before applying the large-\Nc analysis, the tensor interaction would have erroneously been  considered to be \NtwoLO in \Nc.

Before we explore how well the relationships of Eq.~(\ref{NLOrelns}) are satisfied experimentally in Sec.~\ref{sec:LO} we will address some subtleties related to the subtraction point dependence of our results.


\section{Subtraction point dependence}\label{subtraction}

The large-\Nc analysis provides estimates of the relative sizes of \emph{coefficients} in \eftnopi. In general only observables, which are linear combinations of coefficients multiplied by matrix elements, are subtraction point ($\mu$)  independent. Individual EFT coefficients, however, can be $\mu$ dependent, such as \Ctrip, \Csing, and \Csd in our case. 
To compare the large-\Nc predictions, which are relationships among EFT coefficients, to values extracted from experiment, it is therefore necessary to choose a value for $\mu$.
As shown in Ref.~\cite{Kaplan:1995yg} for the LO EFT couplings, some choices can completely hide the additional symmetry emerging in the large-\Nc limit. 
The form of the LO-in-\eftnopi terms can be chosen as (see Refs.~\cite{Weinberg:1990rz,Weinberg:1991um}) 
\begin{equation}
\calL_\text{LO-in-\eftnopi} = -\frac{1}{2} C_S (N^\dagger N) (N^\dagger N)  -\frac{1}{2} C_T (N^\dagger \sigma_i N) (N^\dagger \sigma_i N) \ .
\end{equation}
As discussed in previous sections, this also turns out to be the basis convenient for determining large-\Nc scaling. In the large-\Nc limit $C_S\sim \Nc$, while $C_T\sim 1/\Nc$ \cite{Kaplan:1995yg}. The LECs $C_{S/T}$ are related to the LECs in the partial-wave basis of Eq.~\eqref{eq:LOSwave} by 
\begin{equation}
\label{CSCT}
\Czerosing= (C_S - 3 C_T) \ , \quad \Czerotrip = (C_S + C_T) \ ,
\end{equation}
which suggests that in the large-\Nc limit 
\begin{equation}\label{ratio0}
\left. \frac{\Czerotrip}{\Czerosing}  = \frac{\frac{1}{\asing}-\mu}{\frac{1}{\atrip}-\mu} \right\vert_\text{\LONc} = 1 \ ,
\end{equation}
where we have used the PDS expression of Eq.~\eqref{leadingCs}, with corrections suppressed by a factor of $1/\Nc^2$. If $\mu = 0$ is chosen as the subtraction point, this expression yields a ratio that is not only far from unity but negative.   This is an example of a choice for $\mu$ that violates the large-\Nc prediction. As discussed in Ref.~\cite{Kaplan:1995yg}, this problem is related to the fine-tuning of the scattering lengths and can be avoided by an appropriate choice of the matching scale. In the PDS formalism, this corresponds to choosing a $\mu$ that is larger than the scattering lengths. With increasing $\mu$, the ratio approaches 1 from below, reaching a value of 0.9 (i.e., within $10\%\approx 1/\Nc^2$ of the LO prediction) for $\mu\approx \SI{440}{MeV}$.
However, the 10\% prediction comes from expected corrections to the coefficients in the large-\Nc-counting basis, which in this case are the LECs $C_T$ and $C_S$  rather than the partial-wave coefficients.   
The factor of 3 in Eq.~\eqref{CSCT} modifies the  error estimate on the ratio of Eq.~\eqref{ratio0} to $\approx 4/\Nc^2 \sim 1/\Nc$. 
Given this estimate,  we might expect the ratio in Eq.~\eqref{ratio0} to be no closer than 0.7 (from $1/\Nc\sim 30\%$ corrections), which is obtained at a value of $\mu=\SI{140}{MeV}$.  As we will see below, this is indeed in the range of $\mu$ where the large-\Nc predictions are satisfied for the two-derivative partial-wave operators. On the other hand, at this value of $\mu=\SI{140}{MeV}$, the ratio $C_T/C_S \approx 0.08$, consistent with the expected $1/\Nc^2$ suppression. 

When attempting to relate quantities that have $\mu$ dependence, the aim is to avoid obscuring large-\Nc relationships, as well as to  minimize the $\mu$ dependence and also maintain other features (such as power counting) of the theory. 
In \eftnopi power counting, the large scattering lengths are counted as $1/a \sim Q$, with $Q$ the generic small scale in the theory. Using the PDS scheme, choosing $\mu$ such that $\mu \sim Q$ and simultaneously $(1/a-\mu) \sim Q$ results in LO (in \eftnopi counting) coefficients that justify the resummation of an infinite series of diagrams. 
While the second condition is simple to satisfy in the $\oneS$ channel with a negative scattering length, the positive and smaller scattering length in the $\threeS$ channel requires special care. As can be seen from Eq.~\eqref{leadingCs}, choosing $\mu \approx 1/\atrip$ results in a large denominator. 
The same large denominator problem occurs in the NLO-in-\eftnopi-counting coefficient $\Ctrip$: 
\begin{equation}
\label{NLOCs}
\Ctrip=\frac{2\pi}{M} \frac{\rtrip}{(1/\atrip-\mu)^2}  \  ,
\end{equation}
where $\rtrip$ is the effective range in the spin-triplet channel.
The \eftnopi power counting requires $p^2 \Ctrip$ to be suppressed relative to the LO coefficient $\Czerotrip$. For the large \eftnopi momentum $p\sim \SI{70}{MeV}$, the ratio $p^2 \Ctrip/\Czerotrip$
is less than $1/3$ for $\mu > \SI{100}{MeV}$. Alternatively, one can consider the expansion of $p \cot \delta^{(\threeS)}$ about the deuteron pole \cite{Bethe:1949yr,Bethe:1950jm} instead of about $p=0$ as done above. In that case, the LEC $\Ctrip$ will have an expansion in powers of the \eftnopi parameter $Q/\Lambda_{\text{\eftnopi}}$,
\begin{equation}
\Ctrip = C_{2,-2}^{(^3 \! S_1)} + C_{2,-1}^{(^3 \! S_1)} +\ldots \ ,
\end{equation}
(see Eq.~(2.19) of Ref.~\cite{Chen:1999tn}), where the subscript on the left-hand side and the first subscripts on the right-hand side indicate the number of derivatives of the corresponding operator and the second subscripts on the right-hand side are the scaling of the coefficient with $Q$.
The second term on the right-hand side is suppressed by a factor of $1/3$ relative to the first only for $\mu \gtrsim \SI{100}{MeV}$. We therefore do not expect to be able to take $\mu < \SI{100}{MeV}$ without encountering unnaturally large values of the LECs.
On the other hand, Eq.~\eqref{NLOCs} shows that taking $\mu$ to be very large compared to $1/a$ results in unnaturally small values of the LECs.

The possibility of choosing a renormalization scale of the order of the pion mass is discussed in Ref.~\cite{Epelbaum:2017byx}, whose authors argue that such a choice results in an EFT with LECs that are of natural size, but in which the relative ordering of the perturbative expansion is modified.
Reference~\cite{CalleCordon:2008cz,*CalleCordon:2009ps,*RuizArriola:2016vap} also emphasizes the importance of scale dependence when investigating large-\Nc and related symmetries in NN potential models.

\section{LO-in-large-\Nc Relationships and Comparisons}
\label{sec:LO}

How well do these relationships work in the real world? 
There are seven partial-wave terms, but only three that are leading order in \Nc; the large-\Nc analysis predicts relationships between the different partial-wave couplings.  This is just as anticipated: the number of independent LECs dictated by \eftnopi is reduced when the approximate symmetry of large-\Nc is imposed.

The  \LONc relationships between the LECs in the large-\Nc counting and partial-wave bases can be written in matrix form as 
\begin{equation}\label{pw.2.LO.in.N.matrix}
\def\arraystretch{1.2} 
\begin{pmatrix}
\Csing \\ \Ctrip \\ \Csd \\ \ConeP \\ \CPzero \\ \CPone \\ \CPtwo
\end{pmatrix}
=
\def\arraystretch{1.2} 
\begin{pmatrix}
-4 & 12 &  4 \\
-4 & 12 &  4 \\
0 & 0 & -12 \\
-4 & -36 & -12 \\
-\frac{4}{3} & -\frac{4}{3} & 4 \\
-2 & -2 & -4\\
-1 & -1 & 0 
\end{pmatrix}
\begin{pmatrix}
\couplingOneZero \\ \couplingThreeOne \\\couplingSixOne \end{pmatrix} \ .
\end{equation}
The linear combinations
\begin{equation}
C^D_1 \equiv \couplingOneZero +9 \couplingThreeOne + 3\couplingSixOne \ , \quad  C^D_2 \equiv \couplingOneZero + \couplingThreeOne \ , \quad C^D_3 \equiv \couplingSixOne
\end{equation}
``block-diagonalize'' the $P$ waves,
\begin{equation}
\label{blockdia}
\def\arraystretch{1.2} 
\begin{pmatrix}
\Csing \\ \Ctrip \\ \Csd \\ \ConeP \\ \CPzero \\ \CPone \\ \CPtwo
\end{pmatrix}
=
\def\arraystretch{1.2} 
\begin{pmatrix}
2 & -6 &  -2 \\
2 & -6 &  -2 \\
0 & 0 & -12 \\
-4 & 0 & 0 \\
0 & -\frac{4}{3} & 4 \\
0 & -2 & -4\\
0 & -1 & 0 
\end{pmatrix}
\begin{pmatrix}
C^D_1 \\ C^D_2 \\C^D_3 \end{pmatrix} \ .
\end{equation}
The dependence of the partial-wave-basis coefficients on the leading-order coefficients in the large-\Nc-counting basis can vary by large (compared to $\Nc=3$) factors. For example, there is a relative factor of $9$ between the $\couplingOneZero$ and $\couplingThreeOne$ contributions to the $\oneP$ coupling in Eq.~\eqref{pw.2.LO.in.N.matrix}. These large numerical factors highlight that while the large-\Nc expansion corrections are $1/\Nc^2 = 1/9 \sim 10\%$ at this order, other physics can obscure this counting.  
This is why a dual-expansion (large-$\Nc$ along with \eftnopi) treatment is important, and why large-$\Nc$ predictions alone need to  be treated as trends with potentially substantial errors rather than strict rules for ordering the relative size of observables.

\subsection{$S$ waves}
\begin{figure}
\begin{center}
\includegraphics[width=0.9\textwidth]{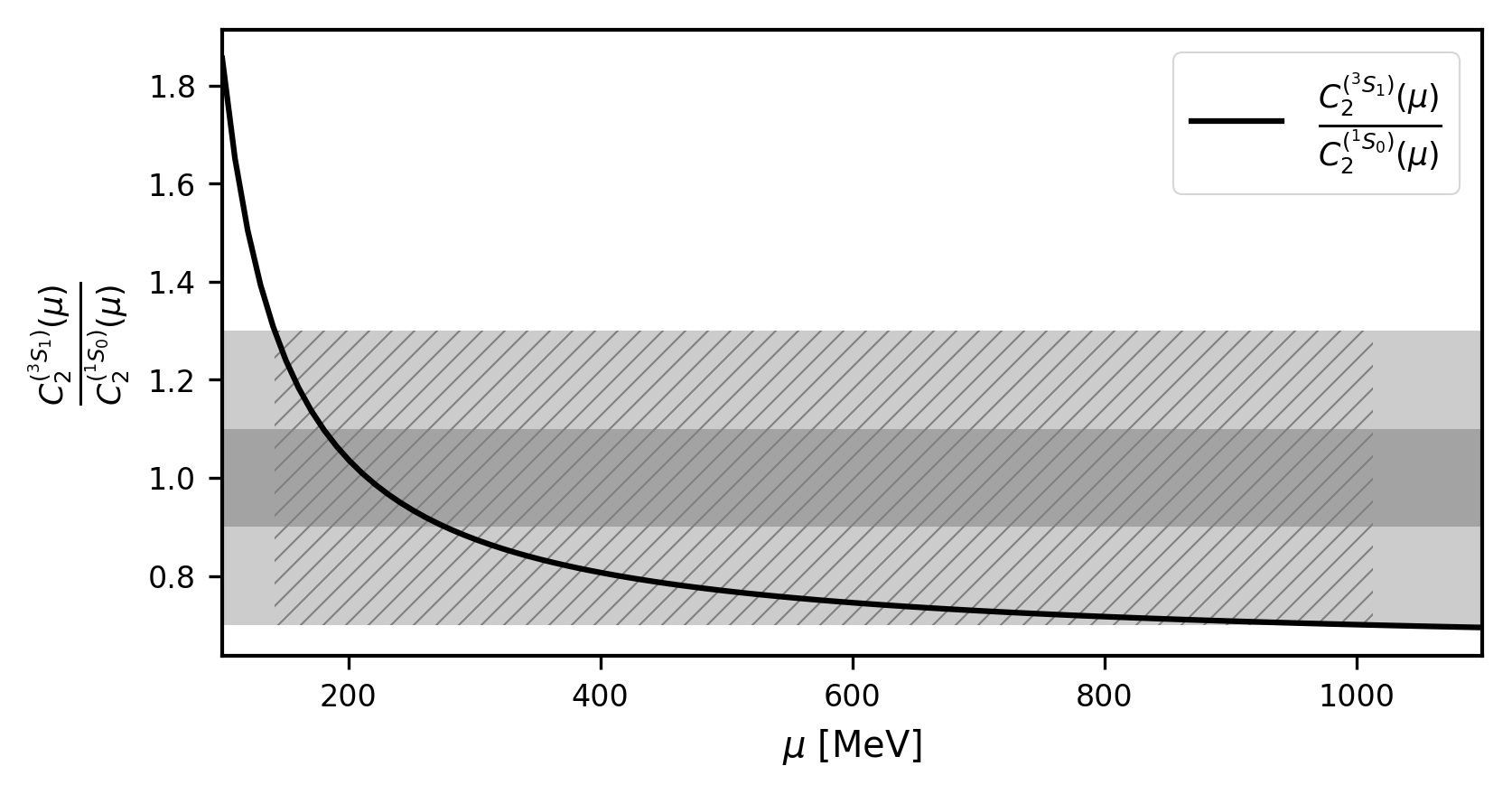}
\end{center}
\caption{The ratio $\Ctrip/\Csing$ of Eq.~\eqref{Ratio1} as a function of the renormalization scale $\mu$ (black curve). The gray band shows the large-\Nc prediction with 10\% (dark gray) and 30\% (light gray) variation. The hatched region indicates the range of $\mu$ over which the large-\Nc prediction is satisfied to 30\%. }
\label{fig-R1}
\end{figure}

As \Nc$\rightarrow \infty$ the prediction from  Eq.~\eqref{pw.2.LO.in.N.matrix} is 
\begin{equation}
\label{Ratio1}
 \left. \frac{\Ctrip}{\Csing}\right|_{\LONc} = 1 \ . 
 \end{equation}
As discussed above, the LECs are not observables and are renormalization-point dependent. 
In the PDS renormalization scheme, the ratio of the $S$-wave couplings is given by
\begin{equation}
  \frac{\Ctrip}{\Csing}  = \frac{\rtrip}{\rsing} \frac{(\mu - 1/\asing)^2}{(\mu - 1/\atrip)^2} \ ,
\end{equation}
where the experimentally extracted values
  \cite{Hackenburg:2006qd} 
\begin{align}
\asing &= \SI{-23.7148 \pm 0.0043}{fm} \ , &
\rsing &= \SI{2.750 \pm 0.018}{fm}\ , \\
\atrip &= \SI{5.4112 \pm 0.0015}{fm}\ , &
\rtrip &= \SI{1.7436 \pm 0.0019}{fm}\ ,
\end{align} 
are the scattering lengths and effective ranges in the $\oneS$ and $\threeS$ channels. 
Figure~\ref{fig-R1} shows the $\mu$ dependence of this ratio. The LO-in-\Nc prediction is satisfied within a 30\% error over a wide range of $\mu$.  
Placing a 10\% or 30\% error bar on the large-\Nc prediction is not meant to be a rigorous analysis of how a 10\% error on the large-\Nc basis coefficients translates to the expected error on the ratio, but is simply an estimate to see if expected trends are satisfied. Although corrections to individual LECs are expected to be $O(1/\Nc^2)\approx 10\%$, the inclusion of the 30\% error on the ratio is motivated by the analogous analysis of the LO-in-\eftnopi results discussed in Sec.~\ref{subtraction}.

This range of $\mu$ agrees with the one found for the LO-in-\eftnopi LECs discussed in Sec.~\ref{subtraction}, which is consistent with the fact that the $\mu$ dependence of the two-derivative LEC in a given $S$-wave channel is related to that of the corresponding zero-derivative LEC through the renormalization group.

\subsection{$P$ waves}\label{Pintro}

The $P$-wave LECs are renormalization scale independent to this order in \eftnopi. Their values can be determined from fits to $P$-wave phase shifts (see Appendix~\ref{app_pwaves} for details). 
For the following discussion we will use the central values 
\begin{equation} \label{Pwave_num}
\begin{alignedat}{3}
  \CPzero & = \SI{6.6}{fm^4} \ , \quad&\CPone & = \SI{-6.0}{fm^4} \ , \\
  \CPtwo & = \SI{0.57}{fm^4} \ , \quad&\ConeP & = \SI{-22}{fm^4} \ .
\end{alignedat}
\end{equation}
These values are in agreement with those of Ref.~\cite{Margaryan:2015rzg}, which were extracted from the Nijmegen potential model at \SI{1}{MeV}. 
Taking into account different energy ranges used in the fits and different models, the potential model extraction of scattering data yields uncertainties on the values of Eq.~\eqref{Pwave_num} of less than 5\%.
Below we will extract values for the large-\Nc basis coefficients to compare with our predictions. The truncation of the large-\Nc expansion at a given order introduces theoretical uncertainties that should be taken into account in the fits. A rigorous treatment of these errors is beyond the scope of this work. Instead, as very rough estimates of the expected theoretical uncertainty of $O(1/\Nc^2)$, we include 10\% errors on the ``experimental" coefficients of Eq.~\eqref{Pwave_num} and apply naive error propagation to estimate errors on the large-\Nc LECs extracted from the fit.
But this should be interpreted with caution and the quoted errors should not be considered rigorous. As usual with large-\Nc predictions it is more appropriate to speak of trends than of predictions with rigorous error estimates.

\subsubsection{${}^3 P _J$ waves}

At \LONc, the ${}^3 P _J$-wave LECs depend upon only two independent parameters; a relationship exists amongst the ${}^3 P _J$-wave LECs.  There are several ways to express this. First, consider  
\begin{equation}\label{Ratio2}
  \left. \frac{\CPzero + \CPone}{\CPtwo} \right|_{\LONc} = \frac{10}{3} \approx 3.3 \ .
\end{equation} 
As can be seen from the numerical values in Eq.~\eqref{Pwave_num}, this is a ratio of small numbers that depends on large cancellations and is therefore very sensitive to the numerical inputs. Using the central values of Eq.~\eqref{Pwave_num} gives a ratio of about 1, apparently in serious disagreement with the large-\Nc prediction.  However, if  the procedure described in Sec.~\ref{Pintro} is used for roughly estimating errors, the result is 0.95 $\pm$ 1.6.  An equivalent alternative is 
\begin{equation}
\label{Ratio2mod}
 \left. \frac{\CPzero-\frac{4}{3}\CPtwo}{-\CPone+2 \CPtwo} \right|_{\LONc} = 1 \ .
\end{equation}
Using the central values of Eq.~\eqref{Pwave_num} gives 0.82.
 Normally distributed uncorrelated 10\% errors about the central values of the partial-wave couplings yield 0.82 $\pm$ 0.12, which is consistent with an \LONc prediction.

\subsubsection{ ${}^1 P _1$ and ${}^3 P _J$ waves} \label{allP}

At LO-in-\Nc, the LEC for the $\oneP$ wave depends on an independent set of couplings compared to the $\PJ$ waves, as  is most easily seen from the lowest four rows of Eq.~\eqref{blockdia}.
In particular,  the $\PJ$-waves in isolation only depend on two independent combinations of LO LECs in the large-\Nc basis.  Including the $\oneP$ requires all three \LONc LECs and it is interesting to discover which values for the large-\Nc-basis LECs at LO are preferred by only the $\mu$-independent partial waves.
Using the last four rows of Eq.~\eqref{pw.2.LO.in.N.matrix} we find that a simultaneous fit of the three LO-in-\Nc LECs to the central values of the ${}^1 P _1$ and ${}^3 P _J$  couplings yields 
\begin{equation}
\couplingOneZero = \SI{-0.36 \pm 0.26}{fm^4} \ ,\qquad  \couplingThreeOne = \SI{0.12 \pm 0.08}{fm^4} \ ,\qquad \couplingSixOne = \SI{1.6 \pm 0.12}{fm^4} \  ,
\end{equation}
where the errors are obtained  by propagating 10\% errors assigned to Eq.~\eqref{Pwave_num}. But such a result is not to be taken too seriously because we have not attempted to fit to all seven partial waves.
In the next two sections we consider large-\Nc relationships that involve the $\mu$-dependent LECs.

\subsection{\SD{}-mixing term and $\PJ$ waves}
\label{sd.ratio}

The \eftnopi LEC for the \SD{}-mixing term is given at its leading order by \cite{Chen:1999tn,Chen:1999vd}
\begin{equation}
{\Csd} = E_1^{(2)} \frac{3}{\sqrt{2}} \Czerotrip  \ ,
\end{equation}
where $E_1^{(2)}$ is  the ($\mu$-independent)  LO coefficient in a momentum expansion of the \SD-mixing parameter $\bar{\epsilon}_1$.\footnote{Because of the slow convergence of the expansion in $E_1^{(n)}$ (where $n$ is the order in the expansion), \Csd is sometimes expressed in terms of $\eta_{SD}$, the asymptotic $D/S$ ratio of the deuteron \cite{Chen:1999vd}.} On dimensional grounds $E_1^{(2)}$ scales as $(1/\LambdaNoPion)^2 \sim \SI{2}{fm^2}$.  But the value extracted from partial-wave data is about five times smaller: $E_1^{(2)} \sim \SI{0.4}{fm^2}$ \cite{Stoks:1994wp,Chen:1999vd}.  This has led to the conclusion that the \SD-mixing coefficient \Csd is unnaturally small.\footnote{We thank G. Rupak for a discussion of this point.}

\begin{figure}
\begin{center}
\includegraphics[width=0.9\textwidth]{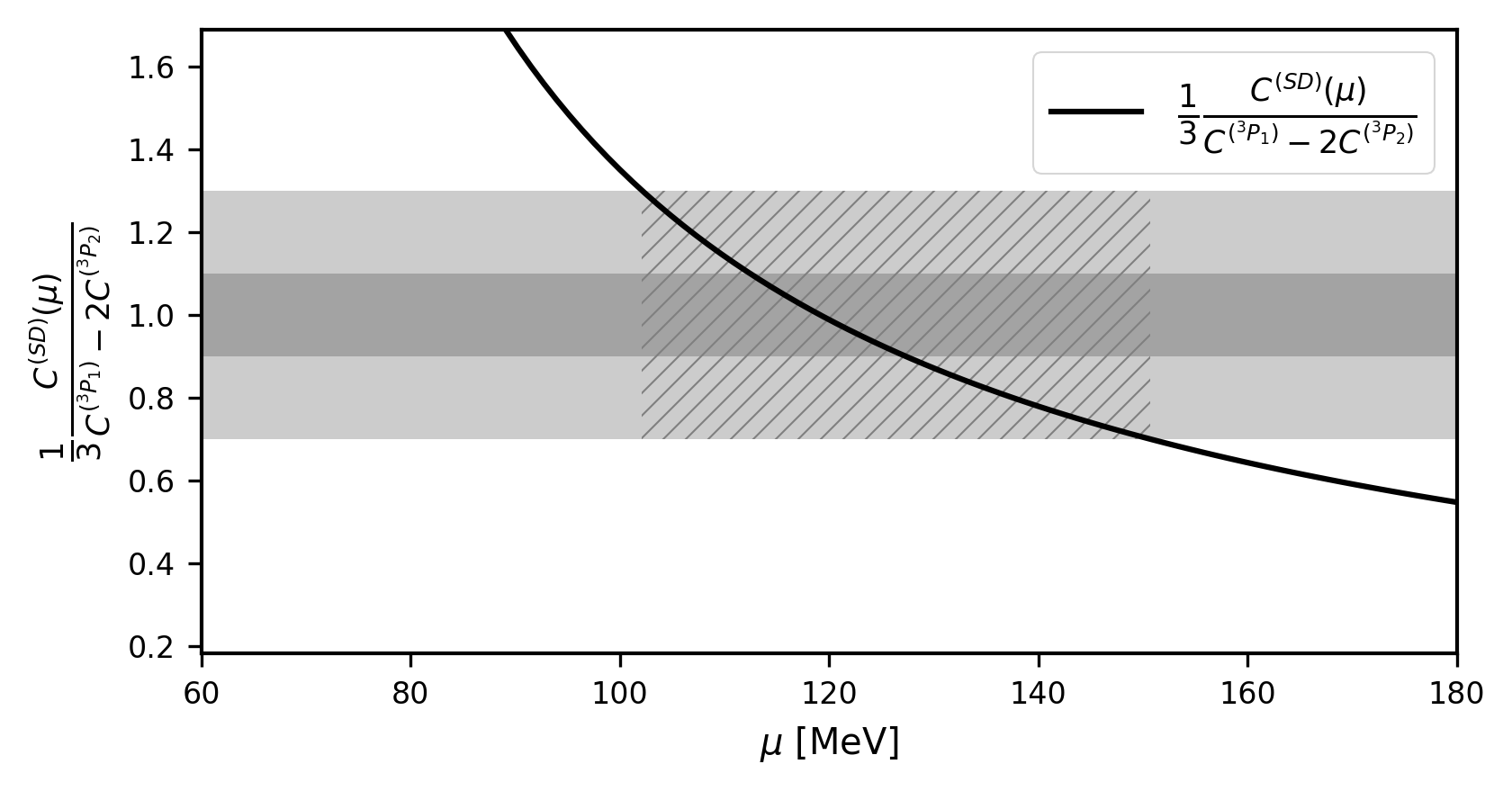}
\end{center}
\caption{The ratio of Eq.~\eqref{Ratio3} as a function of the renormalization scale $\mu$ (black curve) with the value of $\Csd$ a factor of 3 larger than its physical value. The gray band shows the large-\Nc prediction with 10\% (dark gray) and 30\% (light gray) variation. The hatched region indicates the range of $\mu$ over which the large-\Nc prediction is satisfied to 30\%.}
\label{fig-R3-x3}
\end{figure}

A \LONc relationship involving the \SD{}-mixing LEC and ${}^3 P _J$-wave LECs is 
\begin{equation}
\label{Ratio3}
\left. \frac{1}{3} \frac{\Csd}{\CPone-2\CPtwo} \right|_{\LONc} = 1 \  .
\end{equation}
\begin{figure}
\begin{center}
\includegraphics[width=0.9\textwidth]{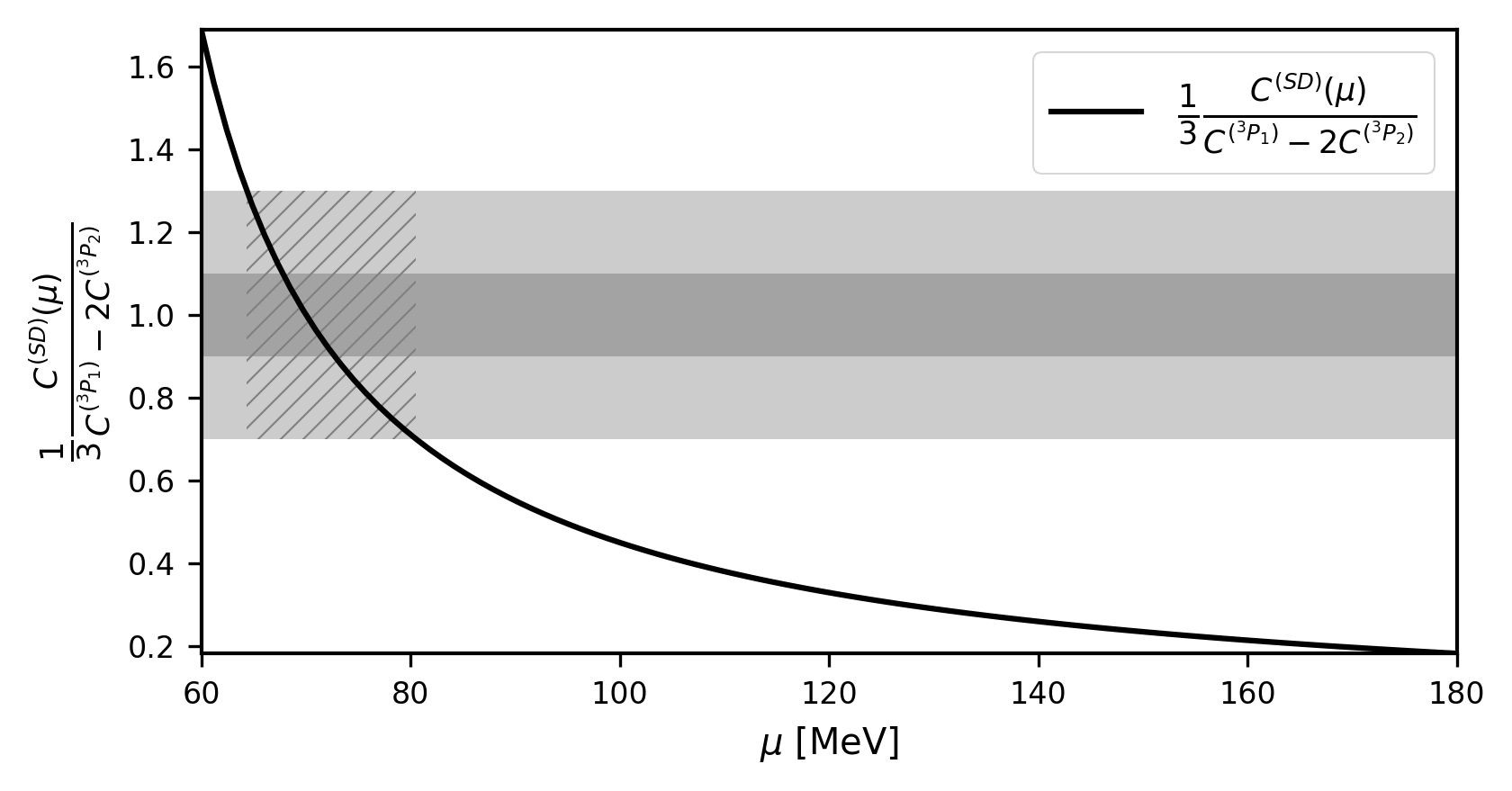}
\end{center}
\caption{The ratio of Eq.~\eqref{Ratio3} as a function of the renormalization scale $\mu$ (black curve). The gray band shows the large-\Nc prediction with 10\% (dark gray) and 30\% (light gray) variation.  The hatched region  indicates the range of $\mu$ over which the large-\Nc prediction is satisfied to 30\%. }
\label{fig-R3}
\end{figure}
Figure~\ref{fig-R3} shows the $\mu$ dependence of this ratio. Agreement with the  \LONc prediction is found at lower values of $\mu$ compared to those found in Fig.~\ref{fig-R1}.  
But  taking into account the suppression of  \Csd  brings the preferred $\mu$ value of  the large-\Nc prediction of Eq.~\eqref{Ratio3} back into alignment with the $\mu$ range preferred by the large-\Nc prediction of Eq.~\eqref{Ratio1}.
For example, with the physical value of $E_1^{(2)}$, matching the large-\Nc expectation for the ratio of Eq.~\eqref{Ratio3} requires $\mu \sim \SI{70}{MeV}$.  Allowing for some of the suppression to be due to variations within natural ranges and taking $\Csd$ to be a factor of 3 (instead of 5) larger than its physical value, Fig.~\ref{fig-R3-x3} shows agreement with the \LONc prediction in the range $\SI{105}{MeV} \lesssim \mu \lesssim \SI{150}{MeV}$. 

%

\subsection{Relationship involving all $L\le 2$}\label{s-sd-p}

\begin{figure}
\begin{center}
\includegraphics[width=0.9\textwidth]{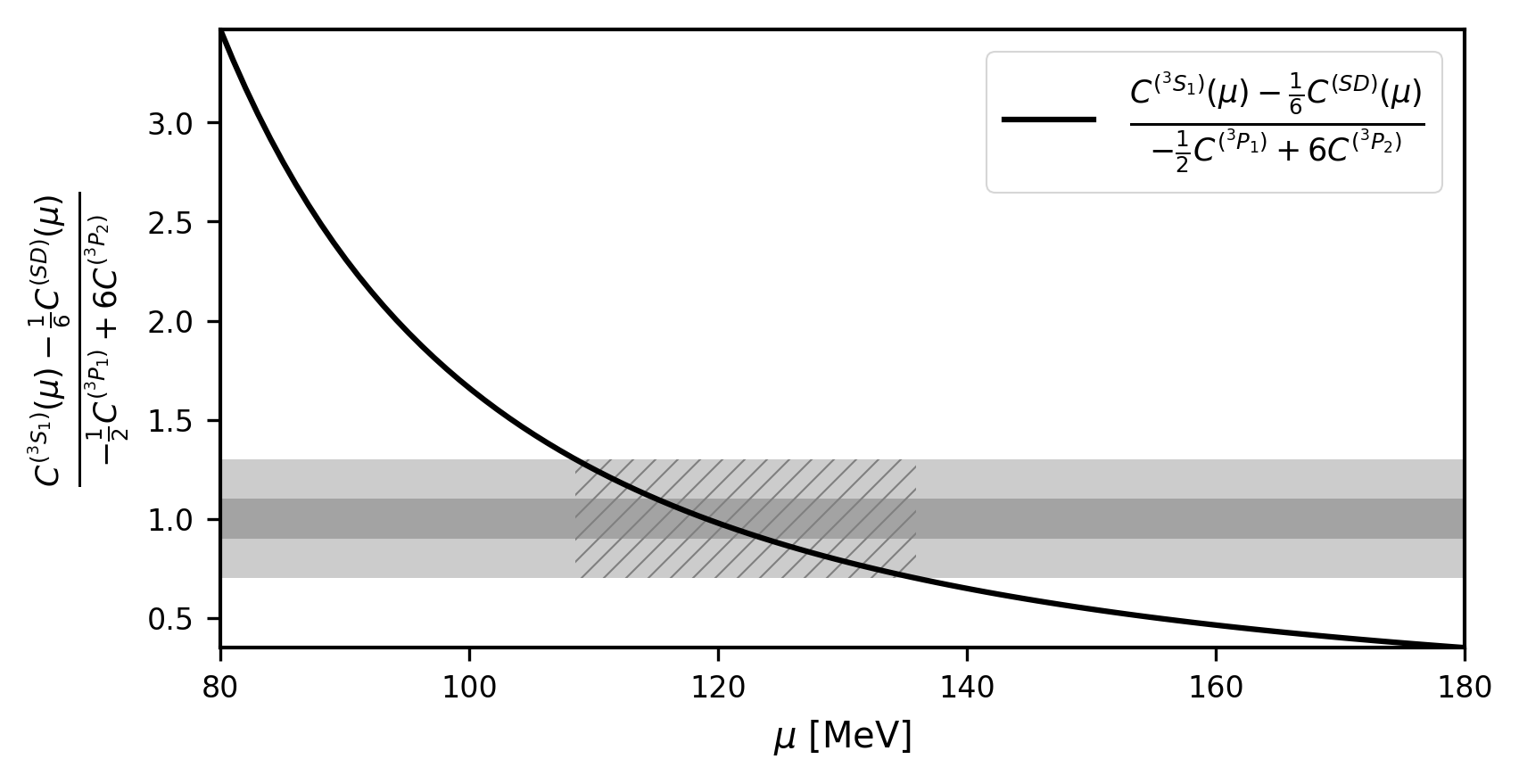}
\end{center}
\caption{The ratio of Eq.~\eqref{Ratio4} as a function of the renormalization scale $\mu$ (black curve). The gray band shows the large-\Nc prediction with 10\% (dark gray) and 30\% (light gray) variation. The hatched region  indicates the range of $\mu$ over which the large-\Nc prediction is satisfied to 30\%.}
\label{fig-R4}
\end{figure}

\begin{figure}
\begin{center}
\includegraphics[width=0.9\textwidth]{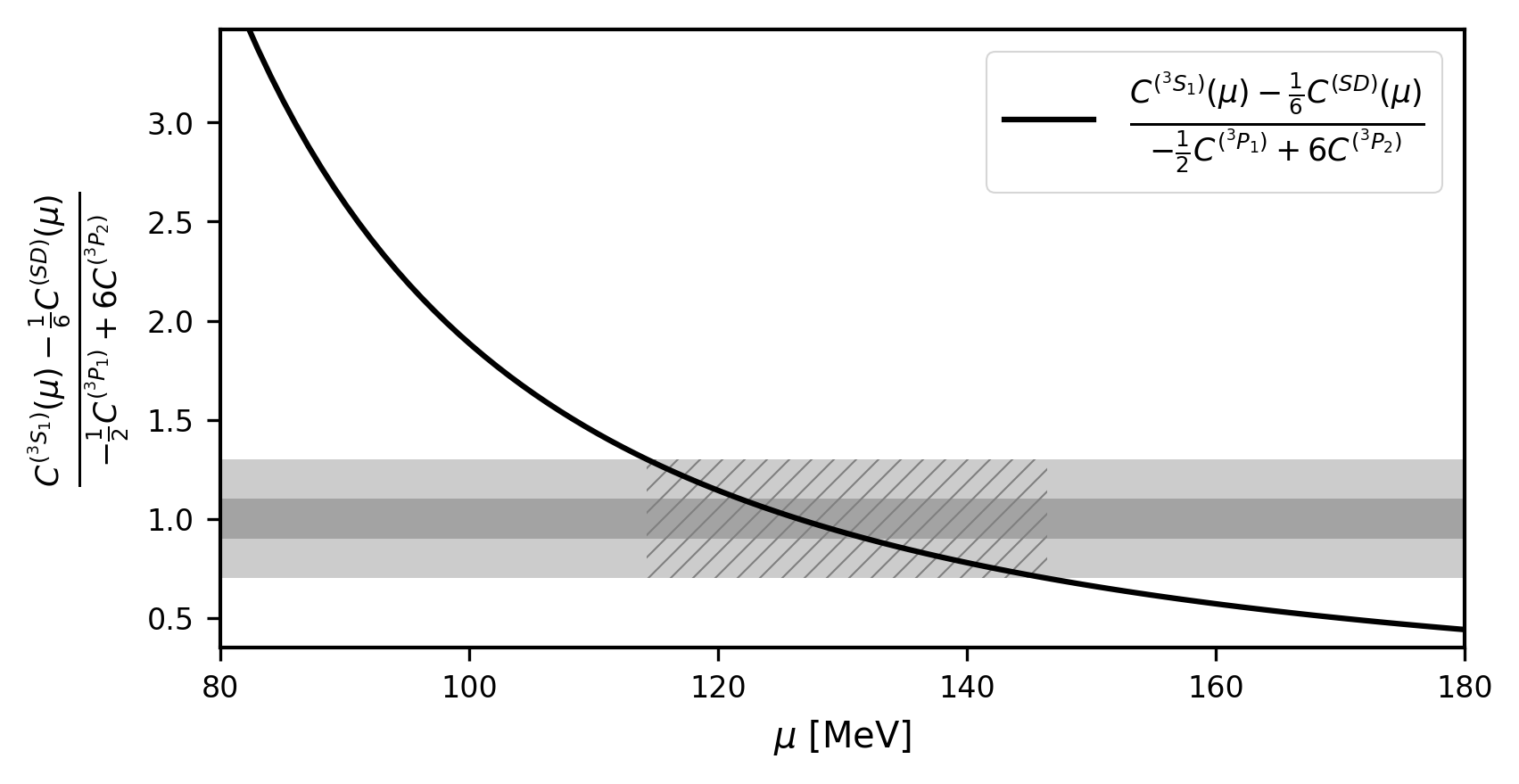}
\end{center}
\caption{The ratio of Eq.~\eqref{Ratio4} as a function of the renormalization scale $\mu$ (black curve) with the value of \Csd a factor of 3 larger than its physical value. The gray band shows the large-\Nc prediction with 10\% (dark gray) and 30\% (light gray) variation. The hatched region indicates the range of $\mu$ over which the large-\Nc prediction is satisfied to 30\%.}
\label{fig-R4-x3}
\end{figure}

The block-diagonal form of Eq.~\eqref{blockdia} shows that at LO in the large-\Nc expansion 
\begin{equation}
\label{Ratio4}
\left. \frac{\Ctrip-\frac{1}{6}\Csd}{-\frac{1}{2} \ConeP +6 \CPtwo}\right|_{\LONc} = 1 \ .
\end{equation}
Figure~\ref{fig-R4} shows this ratio as a function of $\mu$. Agreement within expected 30\% corrections is found for the range $\SI{110}{MeV} \lesssim \mu \lesssim \SI{135}{MeV}$.
A fit of the three LO-in-\Nc coefficients to all seven partial-wave couplings at $\mu = \SI{120}{MeV}$ yields
\begin{equation}
\label{LOfit7}
\couplingOneZero= \SI{-0.58 \pm 0.17}{fm^4} \ ,\qquad \ \couplingThreeOne=\SI{0.42 \pm 0.05}{fm^4} \ ,\qquad \couplingSixOne= \SI{0.76 \pm 0.05}{fm^4} \ ,
\end{equation}
where the errors are estimated by assuming normally distributed uncorrelated 10\% errors about the central values of the partial-wave couplings extracted from experimental inputs for a choice of $\mu=\SI{120}{MeV}$ (see the discussion of error estimates in Sec.~\ref{Pintro}).
While the central values of Eq.~\eqref{LOfit7} do not appear to be very close to those obtained in Sec.~\ref{allP}, these are the more appropriate LECs because the fit involves all partial waves.  Following the discussion of Sec.~\ref{sd.ratio} we now consider how these coefficients change if we allow the anomalously small \Csd coefficient to be three times its experimental value.  In that case, we obtain (still at $\mu=\SI{120}{MeV}$) the following large-\Nc coefficients:
\begin{equation}
\label{LOfit7.3sd}
\couplingOneZero= \SI{-0.59 \pm 0.17}{fm^4},\qquad \couplingThreeOne= \SI{0.10 \pm 0.07}{fm^4},\qquad \couplingSixOne=\SI{1.72 \pm 0.15}{fm^4}.
\end{equation}
Interestingly, these values are more consistent with those found in Sec.~\ref{allP}, indicating that a larger value for \Csd is more compatible with physics in the $P$-wave sector. 
The plot for the large-\Nc ratio of Eq.~\eqref{Ratio4} with \Csd three times its physical value is shown in Fig.~\ref{fig-R4-x3}.  Agreement with the large-\Nc prediction is found for $\SI{115}{MeV} \lesssim \mu \lesssim \SI{145}{MeV}$. 


\section{Results to \NtwoLO in \Nc}

As seen in Sec.~\ref{sec-twoderivative}, four additional independent operators contribute at \NtwoLO in the large-\Nc expansion.
While both central- ($\couplingOneZero, \couplingThreeOne$) and tensor-type ($\couplingSixOne$) interactions are present at \LONc and receive additional \NNLONc contributions ($\couplingOneOne, \couplingThreeZero$, and $\couplingSixZero$, respectively), the spin-orbit contribution ($\couplingFiveZero$) to the ${}^3 P _J$-waves is a new feature at \NNLONc.
 As a result, with seven independent couplings and seven partial waves, the LO relations discussed above no longer hold. For example, as can be seen from Eq.~\eqref{NLOrelns}, the two $S$-wave LECs are no longer predicted to be the same. Instead their ratio takes the form
\begin{equation}
\label{SRatioN2LO}
\left. \frac{\Ctrip}{\Csing} \right\vert_{\NNLONc} =  \frac{\couplingOneZero - 3 \couplingThreeOne - \couplingSixOne  -3 \couplingOneOne + \couplingThreeZero  + \frac{1}{3}\couplingSixZero  }{\couplingOneZero - 3 \couplingThreeOne - \couplingSixOne + \couplingOneOne- 3\couplingThreeZero  - \couplingSixZero } \  .
\end{equation}
The modification of the other ratios considered at LO-in-\Nc is provided in Appendix~\ref{N2LOmod}.

When all seven \NtwoLO-in-\eftnopi LECs are fit to the seven partial-wave LECs at  $\mu=\SI{120}{MeV}$ we obtain 
\begin{align}
  \left.
  \begin{array}{ll}
    \couplingOneZero =\SI{-0.58 \pm 0.11}{fm^4} \ , & \couplingThreeOne = \SI{0.40 \pm 0.05}{fm^4} \ , \\
    \couplingSixOne = \SI{0.84 \pm 0.05}{fm^4}  &
  \end{array} \qquad \right\} &\text{LO}  \\
  \left.
  \begin{array}{ll}
    \couplingOneOne = \SI{0.15 \pm 0.07}{fm^4} \ , & \couplingThreeZero = \SI{-0.39 \pm 0.07}{fm^4} \ , \\
    \couplingSixZero = \SI{0.78 \pm 0.1}{fm^4} \ , & \couplingFiveZero = \SI{-0.17 \pm 0.12}{fm^4} \ ,   
  \end{array} \qquad \right\} &\text{\NtwoLO}
\end{align}
where the errors are again obtained by propagating 10\% errors on the partial-wave LECs.
From this set of values it appears that, at least for this choice of $\mu$, there is no clear evidence that the ``\NNLONc{}'' coefficients are smaller than the ``\LONc{}'' coefficients.  However, all that the large-\Nc analysis tells us is that the \LONc coefficients \emph{can} start at order \Nc, not that they do.  There could be cancellations that cause them to be smaller than expected. It is interesting to see that the spin-orbit term is indeed suppressed compared to the \LONc terms as predicted, although the suppression is less pronounced than the $1/\Nc^2$ expectation.

As discussed in Sec.~\ref{sd.ratio}, obtaining consistent values of $\mu$ to satisfy the large-\Nc ratios tends to confirm the claim of Ref.~\cite{Rupak:1999rk} that the \SD-mixing LEC is unnaturally small. 
Following the approach at LO and taking the \SD-mixing LEC to be a factor of 3 larger than its experimental value yields
\begin{align}
\left.
\begin{array}{ll}
\couplingOneZero = \SI{-0.59 \pm 0.10}{fm^4} \ , & \couplingThreeOne = \SI{0.11 \pm 0.06}{fm^4} \ , \\
 \couplingSixOne = \SI{1.72 \pm 0.13}{fm^4} & 
 \end{array}\qquad \right\} &\text{LO}  \\
 \left.
 \begin{array}{ll}
 \couplingOneOne = \SI{0.16 \pm  0.07}{fm^4} \ , & \couplingThreeZero = \SI{-0.10 \pm 0.08}{fm^4} \ , \\
  \couplingSixZero = \SI{-0.10 \pm 0.16}{fm^4} \ , & 
 \couplingFiveZero = \SI{-0.17 \pm 0.12}{fm^4} \ ,
  \end{array} \qquad \right\} &\text{\NtwoLO}
\end{align}
It is notable that performing the fit of the \NNLONc LECs while excluding the \SD{} term gives very similar results.
These values are closer to the expected pattern that the three LO-in-\Nc LECs  are larger than the N$^2$LO-in-\Nc LECs, with the $\couplingThreeOne$ term the exception. This result emphasizes again that the large-\Nc analysis only provides trends and that other physics can have a significant impact on the size of the couplings.


\section{Conclusions}
\label{sec:conclusions}

We analyzed the two-derivative nucleon-nucleon contact interactions in a combined \eftnopi and large-\Nc approach. The symmetries of QCD as encoded in \eftnopi  restrict the form of the operators, while the corresponding \eftnopi LECs encode short-distance details of the underlying interactions. Because QCD has not been solved, these LECs are treated as free parameters in \eftnopi. The feature of the large-\Nc analysis is that QCD attains additional symmetries in the large-\Nc limit that  constrain the relative sizes of the \eftnopi LECs.  We showed that in the large-\Nc limit only three of the seven \eftnopi LECs are independent; we derived four independent relationships between the LECs that should hold in this limit. 
Critically,  the LECs that multiply operators involving $S$-waves are not observables but contain a subtraction point ($\mu$) dependence.  Since large-\Nc relationships are expressed in terms of these coefficients it is important to choose values of $\mu$ so that the large-\Nc relationships are not obscured. 
By using empirical values extracted from partial-wave analyses, we showed that the \LONc relationships are reasonably well satisfied even in the real world with $\Nc=3$ for appropriate values of $\mu$.  
Consistency among the relations is improved  by implementing the observation that the \SD-mixing coefficient \Csd is unnaturally small and adjusting it upward.

At \LONc the large-\Nc basis coefficients favored by experiment are all of the same order, which suggests that the large-\Nc counting is compatible with nature.  At \NNLONc the values of the three \LONc coefficients do not change dramatically compared to the LO extraction. But contrary to expectation the additional four LECs are not 10\% smaller.  
However, the large-\Nc counting only establishes an upper limit on the size of the \LONc coefficients.
It is possible that other effects cause cancellations that make them smaller than the naive large-\Nc estimate. 
Interestingly, if we take \Csd to be three times its actual value, in acknowledgment that it is experimentally unnaturally small, the three \LONc coefficients again do not change dramatically in the fit involving all seven LECs at \NNLONc, but for this enhanced value of \Csd there is some evidence that the \NNLONc coefficients tend to be smaller compared to the \LONc coefficients.

Our analysis shows that the large-\Nc approach can provide useful guidance in constraining the LECs in \eftnopi. In particular, when little data are available to constrain LECs it may be useful to employ large-\Nc relationships to restrict the number of unknown LECs. However, as illustrated by the \SD-mixing term, other physics that is not captured in the large-\Nc expansion can influence the relative size of couplings. In addition, in the case of subtraction-point-dependent LECs, special care has to be taken to not obscure large-\Nc relations. The results of the large-\Nc analysis should therefore not be considered as exact predictions, but instead should be interpreted as providing overall trends.


\begin{acknowledgments}
We thank G.~Rupak and J.~Vanasse for useful discussions. This material is based upon work supported by the U.S. Department of Energy, Office of Science, Office of Nuclear Physics, under Award Number DE-SC0010300 (M.R.S.) and Award Number DE-FG02-05ER41368 (H.S.~and R.P.S.).
\end{acknowledgments}


\appendix
\section{Fierz identities}
\label{app}
The Fierz identies needed to relate the two bases are:
\begin{align}
\delta_{CB} \delta_{DA} & =  -\frac{1}{2} ( \sigma_2)_{DC} (\sigma_2 )_{AB}+\frac{1}{2}(\sigma_i \sigma_2)_{DC} (\sigma_2 \sigma_i)_{AB} \ ,\\
\delta_{CA} \delta_{DB} & = \frac{1}{2} ( \sigma_2)_{DC} (\sigma_2 )_{AB}+\frac{1}{2}(\sigma_i \sigma_2)_{DC} (\sigma_2 \sigma_i)_{AB} \ , \\
(\sigma_i )_{CB} (\sigma_j )_{DA} &= 
  \frac{1}{2} \delta_{ij} (\sigma_2)_{DC} (\sigma_2)_{AB}   - \frac{1}{2} i \epsilon_{ijk} \left[ (\sigma_k \sigma_2)_{DC} (\sigma_2)_{AB} + (\sigma_2)_{DC} (\sigma_2 \sigma_k)_{AB} \right] \nonumber\\
& \quad - \frac{1}{2} \left( \delta_{ik} \delta_{jn} + \delta_{jk} \delta_{in} - \delta_{ij} \delta_{kn}   \right)
 (\sigma_k \sigma_2)_{DC} (\sigma_2 \sigma_n)_{AB} \ ,  \\
(\sigma_i)_{CA} (\sigma_j )_{DB} &= 
 -\frac{1}{2} \delta_{ij} (\sigma_2)_{DC} (\sigma_2)_{AB} 
 +\frac{1}{2} i \epsilon_{ijk} \left[(\sigma_k \sigma_2)_{DC} (\sigma_2)_{AB} - (\sigma_2)_{DC} (\sigma_2 \sigma_k)_{AB} \right]\nonumber \\
& \quad  - \frac{1}{2} \left(\delta_{ik} \delta_{jn} + \delta_{in}\delta_{jk}  - \delta_{ij} \delta_{kn} \right)
 (\sigma_k \sigma_2)_{DC} (\sigma_2 \sigma_n)_{AB} \ ,
 \end{align}
where  lowercase indices run from 1 through 3 with summation over repeated indices, and uppercase letters run from 1 to 2 as the  spin indices. Analogous relations hold for isospin matrices with the substitution $\boldsymbol{\sigma} \rightarrow \boldsymbol{\tau}$.


\section{Extracting LECs from data}
\label{app_pwaves}

To determine how well the $P$-wave large-\Nc predictions are satisfied we need to  extract the values of the $P$-wave LECs from phase shifts.
They are related using 
\begin{equation}\label{amp}
S^{(c)}=e^{2 i \delta^{(c)}} = 1+i \frac{p M}{2 \pi} A^{(c)}  \ ,
\end{equation}
where  $p$ is the center-of-mass momentum, $M$ is the nucleon mass, $S^{(c)}$ is the scattering $S$ matrix, and $A^{(c)}$ is the scattering amplitude in the $c$ channel. 
The $P$-wave phase shifts $\delta^{(c)}$ can be expanded as~\cite{Chew:1949zz}
\begin{equation}\label{aP}
p^3 {\cot} \ \delta^{(c)} = -\frac{1}{a^{(c)}_P} + \frac{1}{2} r_P^{(c)}p^2 + \cdots  \ ,
\end{equation} 
where $a^{(c)}_P$ is a $P$-wave scattering volume and $r_P^{(c)}$ is a $P$-wave effective range in channel $c$.  For small energies and small phase shifts Eqs.~\eqref{amp} and \eqref{aP} reduce to
\begin{equation}
A^{(c)}=\frac{4 \pi}{p M} \delta^{(c)} + \cdots =\frac{4 \pi a^{(c)}_P}{M} p^2 + O(p^4) \ ,
\end{equation}
where to the order we are considering the amplitudes are related to the LECs by
\begin{equation}\label{alphas}
A^{(\onePone)} = \frac{1}{3}\,p^2\ConeP, \quad A^{(\Pzero)}= p^2 \CPzero, \quad A^{(\Pone)} = \frac{2}{3}\,p^2\CPone, \quad A^{(\Ptwo)}= \frac{4}{3}\,p^2\CPtwo \ .
\end{equation}      
At very low energies, the contribution of partial waves with $L>1$ to a differential cross section can be neglected, but at very low energies the $P$-wave contribution itself is a small percentage of the differential cross section.  Also, low-energy data points are scarce, may have large uncertainties,  and are mostly available from proton-proton scattering, where Coulomb corrections might be important.  Hence there is tension in deciding the most appropriate energy range to use to fit the partial-wave LECs in the $L=1$ sector.  Databases (e.g., NNOnline~\cite{NNOnline} and references therein) helpfully compile available data as encoded via phenomenological potential models. 
The fit here is performed for laboratory energies in the range $[0,T]  \ \si{MeV}$. 
The central values of the fit for $T=\SI{10}{MeV}$ are given in Eq.~\eqref{Pwave_num}.  For  a given potential model the error from using different energy ranges is small.  The variations between different models are about 1\%, with the exception of the $\Ptwo$ wave, for which they are still less than 5\%.
For a detailed study of systematic errors from the extraction of partial wave parameters from potential models see Ref.~\cite{Perez:2014kpa,*Perez:2014waa,*Perez:2016vzj}.

It may also be useful to collect here the relationships relevant for the other two-derivative (but non-$P$-wave) operators. The LECs in the S-wave channels at two derivatives  involve the scattering lengths \asing and \atrip and the effective ranges \rsing and \rtrip.  The relationships with the LECs for channel $S=\oneS$ or $\threeS$ in the PDS scheme are 
\begin{eqnarray}
C_2^{(S)}(\mu)&=&\frac{2 \pi}{M}\left(\frac{1}{-\mu+1/a^{(S)}}\right)^2 r^{(S)} \ \ ,
\end{eqnarray}
 and the connection to the phase shifts is given by the effective range expansion \cite{Schwinger:ERE,Bethe:1949yr,Blatt:1948zz}
 $$p \cot \delta^{(S)} = - \frac{1}{a^{(S)}} + \frac{1}{2} r^{(S)} p^2 + \cdots \ \ . $$ 

The LEC \Csd is related to the \SD-mixing parameter $\bar \epsilon_1$. Performing an expansion of the mixing parameter in powers of $Q$, $\bar \epsilon_1=\epsilon_1^{(2)} + \cdots$, the leading-order contribution appears at order $Q^2$ and is given by \cite{Chen:1999vd}
\begin{equation}
\epsilon_1^{(2)} =  E_1^{(2)}\frac{p^3}{\sqrt{p^2+\gamma^2}} \ ,
\end{equation}
where $E_1^{(2)}=\SI{0.386}{fm^2}$ and $\gamma$ is the deuteron binding momentum. 
The $\mu$-dependent LEC \Csd is given by 
\begin{equation}
\Csd = \frac{3}{\sqrt{2}}  \frac{4\pi}{M}\frac{1}{(\gamma-\mu)} E_1^{(2)} \ .
\end{equation}


\section{\NtwoLO modification of LO large-\Nc ratios}
\label{N2LOmod}

 Analogous to the $S$-wave ratio of Eq.~\eqref{SRatioN2LO}, the other ratios discussed in Sec.~\ref{sec:LO} are also modified by \NtwoLO large-\Nc terms.
The ratio of Eq.~\eqref{Ratio2} takes the form
\begin{equation}
\left. \frac{\CPzero + \CPone}{\CPtwo} \right\vert_{\NNLONc} = 
 \frac{\frac{10}{3} (\couplingOneZero + \couplingThreeOne +\couplingOneOne  + \couplingThreeZero  )- \frac{14}{3} \couplingFiveZero}
 {(\couplingOneZero +  \couplingThreeOne+ \couplingOneOne+ \couplingThreeZero)+  \couplingFiveZero} \ .
\end{equation}
The result is written with suggestive parentheses to indicate that for this large-\Nc expression only the spin-orbit term ($\couplingFiveZero$) modifies the LO-in-\Nc result.   If $\couplingFiveZero$ is smaller than the other $1/\Nc$ terms, this relationship may be considered more robust than those that receive corrections from the central and tensor $1/\Nc$ terms. 
Equation~\eqref{Ratio2mod} is modified to
\begin{equation}
\left. \frac{\CPzero-\frac{4}{3}\CPtwo}{-\CPone+2 \CPtwo} \right\vert_{\NNLONc} = \frac{(\couplingSixOne+\couplingSixZero)+\couplingFiveZero}{(\couplingSixOne+\couplingSixZero)-\couplingFiveZero}    \ .
\end{equation}
Again, since this is an equivalent relationship to the one above, we see that the change in the ratio is only due to the spin-orbit term. If that term is small for some reason then this ratio is less sensitive to \NNLONc corrections. 
From Eq.~\eqref{Ratio3} we find 
\begin{equation}
\left. \frac{1}{3}  \frac{\Csd}{\CPone-2\CPtwo} \right\vert_{\NNLONc} = \frac{1}{3} \ 
 \frac{  3\couplingSixOne - \couplingSixZero}{\couplingSixOne-\couplingFiveZero+ \couplingSixZero} \ ,
\end{equation}
while the relationship of Eq.~\eqref{Ratio4} takes the form
\begin{align}
&\left. \frac{\Ctrip-\frac{1}{6}\Csd}{-\frac{1}{2} \ConeP +6 \CPtwo} \right\vert_{\NNLONc} = \nonumber \\ 
& \frac{ -2 \couplingOneZero +6\couplingThreeOne +3 \couplingSixOne  +6 \couplingOneOne -2 \couplingThreeZero  - \couplingSixZero}
{    -2 \couplingOneZero +6 \couplingThreeOne +3 \couplingSixOne  -6 \couplingOneOne -6\couplingThreeZero  - \couplingSixZero  
     - 3\couplingFiveZero    }  \ .
 \end{align}



%

\end{document}